\documentstyle[12pt]{article}
\begin{document}
\voffset= -0.5cm
\hoffset= -0.5cm
\textwidth= 18cm
\textheight= 20cm
\baselineskip 18pt
\author{ M.A. Alpar\\
Physics Department, Middle East Technical University\\
Ankara 06531, Turkey\\
Present address: Sabanc{\i} University\\
Istanbul 81474, Turkey }
\title{Isolated thermal neutron stars, soft gamma-ray repeaters and
anomalous X-ray pulsars: propellers and early accretors with
conventional magnetic fields?}
\maketitle
\noindent
%\begin{abstract}
%\vspace{3cm}
{\bf  
The similarity of rotation periods from three interesting classes of 
neutron stars, the anomalous X-ray pulsars (AXPs), the soft gamma ray 
repeaters (SGRs) and the dim isolated thermal neutron stars (DTNs) suggests a 
common mechanism with an asymptotic spindown
phase, extending through the propeller and early accretion stages. 
The DTNs are interpreted as sources in the propeller stage. 
Their low luminosities arise from frictional heating in the neutron star. 
SGRs and AXPs are accreting at $\dot{M} \sim 10^{15} gm/s $.  
The limited range of near equilibrium periods corresponds to a limited 
range of mass inflow rates $\dot{M}$. For lower rates the source of 
mass inflow may be depleted before the asymptotic stage is reached, 
while sources with higher $\dot{M}$ or later ages possess 
circumstellar material that is optically thick to electron scattering, 
destroying the X-ray beaming and the modulation at the rotation period. 
The model works with conventional magnetic fields of 10$^{11}$-10$^{12}$ G, 
obviating the need to postulate magnetars.  
Frequently sampled timing observations of AXPs, SGRs and DTNs can
distinguish between this explanation and the magnetar model.} 
%\end{abstract}
\newpage  
Anomalous x-ray pulsars (AXPs)$^{1-16}$ are characterized by periods in 
the range 6-12 secs. 
RXJ0720.4-3125, one of the two (or a few) dim nearby ROSAT point sources 
thought 
to be thermally emitting neutron stars$^{17-23}$, has a measured period$^{19}$ 
of 8.4 s, while two of the four confirmed soft gamma ray 
repeaters (SGRs) also have periods in the AXP period range, with 
P = 7.47s for SGR 1806-20 and P = 5.16 s for SGR 1900+14$^{24-29}$. 
For the AXPs it was pointed out$^{1,2}$ that the similar periods would obtain 
as equilibrium periods for accretion rates $\dot{M} \sim 10^{15} gm/s $ 
corresponding to the luminosities and 10$^{11}$-10$^{12}$
G magnetic fields. The limited range of $\dot{M}$ then needs to be
explained$^3$. $\dot{P}$ values measured 
from the AXPs and SGRs, together with the long rotation periods in the 
5-12 s range have led to the suggestion that these sources are 
magnetars$^{29-33}$, neutron stars with very strong magnetic fields
B$\sim $ 10$^{14}$ G spinning down through magnetic dipole radiation. 
The short cooling age estimated for 
RXJ0720.4-3125 has also been suggested$^{34}$ as evidence for a 
magnetar if the source is an isolated pulsar born with a short 
rotation period. We propose here an explanation for all three classes of 
neutron stars with conventional 10$^{12}$ G magnetic fields and rotation 
periods of the order of 10 ms at birth. Presence of mass transfer at 
rates $\dot{M} = 10^{15} - 10^{16}$ gm/s either from a very 
low mass companion$^1$ or a remnant disk$^{2,3}$ 
around the neutron star will produce torques on the neutron star 
of the order of the spindown torques observed. 
The SGRs and AXPs would be accreting at least part of the incoming mass flow. 
The similarity of the periods would simply reflect that all these systems 
are asymptotically approaching equilibrium periods in a common range 
defined by the common ranges of magnetic fields and mass transfer 
rates. As the approach to rotational
equilibrium is asymptotic, $P/\dot{P}$ is not the true age of these 
systems. Sources of different ages and different circumstances have
similar periods in the asymptotic regime if they have 
similar magnetic fields and are subject to spindown under similar mass inflow 
rates. 

All AXPs, SGRs and DTNs are listed in Table 1 together with 
measured pulse periods and $\dot{\em P}$ values. The two DTNs 
yield fits to blackbody spectra with temperatures of 57 eV for 
RXJ185635-3754 and 79 eV for RXJ0720.4-3125 and luminosities 
in the $L_x \sim 10^{31-32}$ erg s$^{-1}$ range$^{17-21}$. 
A few further candidates for this class, which recently emerged from
ROSAT surveys, also have similar blackbody temperatures, flux values and
limits on the ratio of X-ray flux to optical flux$^{22,23}$. 
Accretion from the interstellar medium 
would require unlikely ambient interstellar medium densities and low 
velocities. The cooling of a young 
neutron star can typically yield the observed thermal luminosities at ages 
of the order of 10$^5$-10$^6$ yrs$^{35-37}$. 
For a neutron star born with a rotation period of the 
order of 10 ms, (as typically inferred from the P and $\dot{P}$ values of 
young pulsars), to have spun down to the 8.4 s period of RXJ0720.4-3125 
in 10$^6$ yrs as a rotating magnetic dipole would require a mean spindown 
rate $\dot\Omega$ of the order of 10$^{-14}$ rad s$^{-2}$, 
and a magnetic field of the order of 3 $\times$ 10$^{13}$ G or more$^{34}$. 
The presence of two dim thermal neutron stars within $\sim$ 100 pc 
suggests that there are $\sim$ 10$^4$ such sources in the galaxy,
requiring ages of $\sim $10$^6$ years or longer, if the birth rate is 
10$^{-2}$ yr$^{-1}$ or less.

{\bf Luminosity of a non-accreting neutron star from energy dissipation}

 There is an alternative source of the thermal 
luminosity which takes over at $\sim$ 10$^5$-10$^6$ yrs, after the initial 
cooling, and lasts longer than the cooling luminosity: There will be energy
dissipation (frictional heating) in a neutron star being 
spun down by some external torque. The rate of energy 
dissipation is given by$^{38,39}$ 
\begin{equation}
\dot{E}_{diss} = I_p \omega |{\dot\Omega}|
\end{equation}
where I$_p$ is the moment of inertia of some component of the 
neutron star whose rotation rate is faster than that of 
the observed crust by the amount $\omega$. 
$\dot{E}_{diss}$ will supply the thermal luminosity of a 
non-accreting neutron star at ages greater than $\sim$ 10$^6$ years$^{40-42}$, 
as the cooling luminosity rapidly falls below $\dot{E}_{diss}$ 
after the transition from neutrino cooling to surface photon cooling. 
Among the radio pulsars with X-ray emission$^{35,42}$, 
observations$^{43,44}$ of the pulsar PSR 1929+10, 
whose spin-down age is 3.1 $\times$ 10$^6$ yrs, provide 
an upper limit to the thermal luminosity which yields$^{45}$ 
I$_p \omega <$ 10$^{43}$ gm cm$^2$ rad s$^{-1}$. 

A lower limit to $\dot{E}_{diss}$ can be obtained from 
the parameters of large pulsar glitches$^{45}$. The consistency of the 
observed glitch parameters of all pulsars with large glitches 
and measured second derivatives $\ddot\Omega$ of the rotation rate$^{46}$ 
as well as the statistics of the large glitches$^{47}$ support the 
hypothesis 
that large glitches are a universal feature of pulsar dynamics. 
The current phenomenological description$^{45,48}$ of large glitches entails 
angular momentum exchange between the crust and an interior 
component (a pinned superfluid in current models) which rotates 
faster than the crust by the lag $\omega$. The typical glitch 
related change in relative rotation rate of the crust and interior, 
$\delta\omega \sim$ 10$^{-2}$ rad s$^{-2}$, inferred from the common 
behaviour of all pulsars with large glitches, must be less than the 
lag $\omega$. Using values of I$_p \sim$ 10$^{43}$ 
gm cm$^2$ inferred from the detailed 
postglitch timing measurements available for the Vela pulsar glitches$^{48}$ 
we obtain the lower bound I$_p \omega > $I$_p \delta\omega $= 

10$^{41}$ gm cm$^2$ rad s$^{-1}$. 
This lower bound is independent of the 
details of the glitch models and rests only on the assumption that the 
large glitches involve angular momentum exchange within the 
neutron star. Further, as the 
mode of angular momentum transfer inside the star depends only on 
neutron star structure, the same parameter I$_p \omega$ would 
determine the energy dissipation rates in all neutron stars 
under external torques, also when the source of the external torque 
is not magnetic dipole radiation. The upper and lower bounds together 
imply 
\begin{equation}
10^{41} |\dot\Omega| {\em erg s}^{-1} < L = \dot{E}_{diss} 
< 10^{43} |\dot\Omega| {\em erg s}^{-1} .
\end{equation}
The luminosities of RXJ0720.4-3125 and RXJ185635-3754 give  
\begin{equation}
|\dot\Omega| \sim 10^{-12} - 10^{-10} {\em rad s}^{-2}. 
\end{equation}
With the 8.4s period of RXJ0720.4-3125 
these spindown rates would imply surface magnetic fields in excess 
of 10$^{14}$ G if magnetic dipole spindown is
assumed. While there is no direct observational evidence for 
10$^{14}$ G magnetic fields, 10$^{12}$ G fields are typical of the 
canonical radio pulsars and of the accreting neutron stars with 
observed cyclotron lines.

{\bf Propeller spindown rates}

 Can high spindown rates,  
larger than 10$^{-12}$ rad s$^{-2}$, obtain for neutron stars 
with conventional 10$^{12}$ G fields? This can indeed be expected  
under the spindown torques in certain phases of accreting sources.
For the AXPs this is a 
possibility that has already been explored$^{1-3}$ and will be pursued 
below. In connection with the DTNs, we note  
that a neutron star subject to mass inflow will 
experience high spindown rates even when the inflowing mass is not 
accreted, because of the star's centrifugal barrier (the propeller 
effect)$^{49}$ and propose that RXJ 0720.4-3125 and RXJ185635-3754 are 
neutron stars with magnetic fields of the order of 10$^{12}$ G, spinning
down under propeller torques. The luminosities are produced by
energy dissipation in the neutron star, for just the range of propeller
spindown torques expected for the mass inflow rates indicated by the
luminosities of the accreting AXPs and SGRs as mass accretion rates. 
A neutron star interacting with a mass inflow onto its magnetosphere will not 
accrete if its rotation rate is fast enough to set up an efficient centrifugal 
barrier to the incoming mass$^{49}$. For 
neutron stars born with rapid rotation rates in an ambience with mass 
inflow, a phase of propeller spindown should precede the start of accretion. 
Let us explore the possibility that the DTNs are in the propeller stage
and the AXPs and SGRs in the subsequent accretion phase. 
Sources in the propeller phase have not been detected previously. This is 
understandable since they are not lit up with an accretion luminosity. 
RXJ 0720.4-3125 and RXJ185635-3754, if they are the first observed examples 
of propellers, are observed through their dissipation luminosities, 
only as the nearest members of the DTN class, and only by ROSAT, 
in view of their surface temperatures in the soft X-ray band. 
In the  propeller phase there is a spindown torque 
exerted on the neutron star, through the interaction of its magnetosphere 
with the ambient material. The order of magnitude of this spindown torque is 
\begin{equation}
 N \sim \mu^2 / {r_A}^3 
\end{equation}
where $\mu = B R^3 $ is the magnetic moment of the neutron star 
with surface dipole magnetic field B and radius R, and 
\begin{equation}
r_A = 9.85 \times 10^8 cm {\mu_{30}}^{4/7} 
{{\dot{M}}_{15}}^{-2/7} {\em m}^{-1/7}
\end{equation}
is the Alfven radius$^{50,51}$. Here $\mu_{30}$ denotes the magnetic moment in 
units of 10$^{30}$ G cm$^3$, ${\dot{M}}_{15}$ is the mass inflow rate 
in units of 10$^{15}$ gm s$^{-1}$ and m is the neutron star mass in 
solar mass units. The propeller phase will last 
until accretion starts when a critical rotation period is reached. 
This critical period is of the 
order of, but somewhat smaller than the equilibrium rotation period 
\begin{equation}
P_{eq} = 16.8 {\em s} {\mu_{30}}^{6/7} 
{{\dot{M}}_{15}}^{-3/7} {\em m}^{-5/7}
\end{equation}
which obtains when the star's rotation rate equals the Keplerean 
rotation rate of ambient matter at the Alfven radius; that is, 
when the corotation radius r$_c$ = (GM)$^{1/3} \Omega^{-2/3}$ becomes 
equal to r$_A$ . The neutron star will continue to spin down, now as an 
accreting source, as its period evolves from the critical period towards 
the equilibrium period. Similarity of the rotation periods of the AXPs is 
taken to indicate their proximity to rotational equilibrium at similar 
equilibrium periods. The approach to rotational equilibrium is likely to 
be an asymptotic process extending through the propeller phase on to the 
accretion stage. This provides a natural explanation of the similarity of the 
rotation periods of all AXPs and of the rotation period of 
RXJ0720.4-3125. All these neutron stars could have been born with short 
periods, and similar magnetic moments and encounter similar ranges of mass 
inflow rates. 
Surrounding the neutron star initially there may be material from the debris 
of the core collapse in a supernova explosion or of a Thorne-Zytkow object 
already within the gravitational capture radius of the neutron star, possibly 
in the form of a disk$^{2,3}$. The propeller phase can also occur in the
evolution of a neutron star in a binary, preceding the accretion phase.
In the propeller phase the source settles to an asymptotic spindown towards 
the equilibrium. Accretion starts during the asymptotic spindown phase. A    
source is most likely to be observed during its asymptotic phase. 

>From Eq. (6) we obtain 
\begin{equation}
{\mu_{30}} \sim 0.45 (P_{eq}/8.4 s)^{7/6} {{\dot{M}}_{15}}^{1/2} 
{\em m}^{5/6}  
\end{equation}   
by normalizing P$_{eq}$ to the 8.4 s rotation period of 
RXJ0720.4-3125. 
This estimate of the magnetic moment also applies to the DTNs as
propellers and to AXPs and SGRs as accretors 
asymptotically close to rotational equilibrium. The similarity in
periods ascribed to the asymptotic range then translates, for magnetic
fields in the conventional 10$^{12}$ G range ($\mu_{30} \sim 1$), into a
restricted range of $\dot{M} \sim 10^{15}$ gm s$^{-1}$, so the task is
to understand why a restricted range of $\dot{M}$ prevails in the
observed sources$^{3}$. As accreting neutron
stars the luminosities of the SGRs and AXPs 
indicate accretion rates in the 10$^{15}$ gm s$^{-1}$ 
range. For the non-accreting DTNs  $\dot{M} \sim 10^{15}$ gm s$^{-1}$ is
the rate of mass inflow onto the propeller. This mass inflow, while not
accreting, causes the neutron star to spin down. The propeller torques 
estimated with Eqs.(4) and (5) for conventional magnetic moments 
and with $\dot{M} \sim 10^{15}$ gm s$^{-1}$ produce the spindown rates
inferred for the DTN sources through the interpretation of their
luminosities as due to energy dissipation (Eqs.(1)-(3)). Observed spindown 
rates for the AXPs and SGRs also agree with the rough estimates using
Eq.(4) to order of magnitude.

{\bf Asymptotic spindown}

 We now turn to a simple model for the asymptotic spindown. A neutron star 
in the presence of inflowing matter experiences 
both spindown and spin-up torques. The long term evolution is 
determined by the balance between these, described by a function 
$n(\Omega/\Omega_{eq} - \omega_c)^{52,53}$ which goes through zero when
$\Omega = \omega_c \Omega_{eq} $ where  $\omega_c$ is of order one, and
$\Omega_{eq} = 2\pi /{\em P}_{eq}$. 
In accretion from a disk, the relative specific angular momentum 
brought in by the accreting 
material to spin the neutron star up is 
$[(GMr_A)^{1/2} - \Omega {r_A}^2]$. Since the dimensional torque 
is $ \mu^2 / {r_A}^3 \sim \dot{M} (GMr_A)^{1/2} $ , 
spindown near equilibrium can be modelled as 
\begin{equation}
\dot\Omega = (\mu^2 / I {r_A}^3) ( 1 - \Omega/\Omega_{eq}) 
            = ( \Omega_{eq} - \Omega ) / t_0
\end{equation}
where t$_0$ = $\Omega_{eq} I {r_A}^3 / \mu^2 $.
Here the zero of the torque (the end of the spindown era) is taken to be
at $\Omega = \Omega_{eq}$ rather than $\omega_c \Omega_{eq} $, for
simplicity. The full torque expression throughout the neutron star's history is 
probably more complicated. For example, the torque may contain a factor
$(r_c/r_A)^\beta$ to describe a reduction in the early, rapid 
propeller phases when the star is rotating much faster than the disk 
and $r_c << r_A$; $\beta = 3/2$ is proposed for a supersonic 
propeller$^{50}$; or the magnetic moment could decay 
at a rate proportional to the spindown rate$^{54}$.
These effects lead to initial power law decays of $\Omega$ as a function 
of time, but the final evolution is asymptotic. 
Whatever the form of the initial spindown may be, once the rotation 
rate is close to the equilibrium value, the factor 
$( \Omega_{eq} - \Omega )$ dominates the asymptotic evolution, which 
becomes an exponential decay in time. 
Here we employ the simple asymptotic model of Eq.(11) assuming that the  
magnetic moment is constant, and that the mass inflow rate 
is also a constant representing the long term average $\dot{M}$. 
With constant $\mu$ and $r_A$ , the spindown leads $\Omega$ 
towards $\Omega_{eq}$: 
\begin{equation}
\Omega (t) = [\Omega (0) -\Omega_{eq}] exp ( -t / t_0) +  \Omega_{eq}. 
\end{equation}
This solution is to describe the spindown through both the propeller and 
the accretion phases, 
with mass accretion starting at some $\Omega_{acc} < \Omega_{eq}$. 
>From Eqs.(5),(6) and (12) we obtain
\begin{equation} 
t_0 = 1.3 \times 10^{12} s I_{45} 
{{\dot{M}}_{15}}^{-1} m^{-2/3} {\Omega_{eq}}^{4/3} .
\end{equation}
Substituting this expression in the spindown equation, Eq.(12), 
$\Omega_{eq}$ can be obtained for each source with known $\Omega$ and 
$\dot\Omega$ by solving 
\begin{equation} 
\Omega_{eq} = \Omega + 1.3 \times 10^{12} s \dot\Omega  I_{45} 
{{\dot{M}}_{15}}^{-1} m^{-2/3} {\Omega_{eq}}^{4/3} .
\end{equation}
For the AXPs the mass inflow rate can be equal 
to or somewhat larger than the mass accretion rate inferred from the 
luminosity, as the propeller mechanism may still be partially 
effective. The typical value $\dot{M} \sim $ 10$^{15}$ gm/s of the mass 
inflow rate inferred from the accretion luminosities as the AXPs will 
be adopted here for non-accreting sources like RXJ0720.4-3125. 
Once the equilibrium rotation rate is estimated, the 
magnetic moment $\mu$ can be obtained from Eq.(6), 
\begin{equation} 
\mu_{30} = 0.32 {{\dot{M}}_{15}}^{1/2} m^{5/6} {\Omega_{eq}}^{-7/6} .
\end{equation}
The time since the beginning of the propeller phase can be obtained 
from Eq.(12) for a given  
initial rotation period, which we take to be P(0) = 0.01 s for a newborn
neutron star subjected to propeller spindown 
The time t estimated from Eq.(12) is: 
\begin{equation}
t_{min} = t_0 {\em log} [(\Omega(0) - \Omega_{eq})/ (\Omega - \Omega_{eq})].
\end{equation}
This estimated time t$_{min}$ is actually a lower bound for the time
spent in the propeller and accretion phases. Fluctuations in
$\dot{\em M}$ will erratically offset the asymptotic relaxation by
inducing changes in the "target" $\Omega_{eq}$. 
Then the asymptotic spindown will extend under relatively small short
term changes in $\dot{M}$ as long as $\Omega$ remains larger than
$\Omega_{eq}(\dot{M})$. Each change in $\dot{M}$ would reset the
asymptotic spindown, towards the new $\Omega_{eq}(\dot{M})$.
Thus the asymptotic spindown phase could last much longer than t$_{min}$, 
which is 
a valid estimate for a single uninterrupted value of $\dot{M}$. The actual
duration of the asymptotic phase is likely to be determined by the 
timescale on which $\dot{M}$ becomes large enough to
attain $\Omega_{eq}(\dot{M}) > \Omega$, on the average, so that the 
system enters a sustained spin-up phase.

{\bf Evolution: sources with low $\dot{M}$ may not survive the propeller phase}

Sample solutions for t$_0$ and t$_{min}$ are given in Table 1 
for the AXPs and SGRs with measured P and $\dot{\em P}$ 
and for the dim source RXJ0720.4-3125. 
The AXPs show changes in spindown rate, by up to an order of magnitude, 
on timescales of several years. Long term average values of the 
spindown rate are used, as appropriate for the model to describe the 
long term average trend. Mass inflow rates are inferred from the observed 
luminosities taken as accretion luminosities onto neutron stars. 
In the case of 1E2259+586 a mass inflow rate $\dot{M} \sim $ 10$^{17}$ gm/s, 
rather than the accretion rate $\dot{M} \sim $ 2 $\times$ 10$^{15}$ gm/s must 
be taken to obtain an age estimate t$_{min}$ in agreement with the age estimate 
for the possibly associated supernova remnant CTB 109. For RXJ0720.43125, 
a mass inflow rate of 10$^{15}$ gm/s typical of the AXPs was adopted
since this source is presumably a non accreting, propeller source. Along
with its observed P = 8.47s, $\dot{\Omega} =$ 2.6 $\times $10$^{-12}$ 
($\dot{\em P}$ = 2.9 $\times$ 10$^{-11}$ s s$^{-1}$, comparable to the
AXP values) was inferred from Eq.(1), taking its luminosity as due to
energy dissipation in the neutron star, L = $\dot{\em E}_{diss}$. 
The inferred magnetic fields are all in 
the 10$^{11}$ - 10$^{12}$ G range. The solutions show that ages can be 
much larger than the timescale 
t$_0$ which would give an estimate close to the real age if the spindown 
followed a power law in time rather than an asymptotic relaxation. 
At constant $\dot{M}$, the asymptotic regime would extend indefinitely. 
In practice the duration of the spindown phase is determined by the 
evolution of $\dot{M}$. If all neutron stars are born with similar
magnetic fields, it is the
mass inflow rate that determines the equilibrium period towards which
the spindown leads. Two alternatives have been suggested for the supply
of the mass inflow $^{1,2}$. (i) If $\dot{M}$ is supplied by a remnant disk, 
formed from the debris of the supernova explosion or the progenitor 
evolution of a Thorne-Zytkow object$^{2}$, the spindown phase would be 
terminated by the depletion of the supply of $\dot{M}$. Two out of six
AXPs provide direct evidence of youth through their likely
supernova associations. Here we venture the suggestion that a remnant
disk around the neutron star might also be formed from the debris of the
core regions in a supernova explosion. If all supernovae left neutron stars 
that went through an AXP phase under mass inflow from the debris, then we
would expect at least 100 such objects for a lifetime $>$ 10$^4$ yrs and 
galactic supernova rate of 10$^{-2}$, since the AXPs would be observable
from all galactic distances. That we only observe a few indicates that
mass inflow at the required rates from remnant disks or debris is 
rare, or that such sources rarely reach the accretion phase. The number of 
DTNs in the galaxy must be much larger, 
since we see two within a distance of the order of 100pc, limited by the 
low luminosity of these sources. In the galaxy 
we would expect about 10$^4$ DTNs, and a lifetime of the order of 10$^6$ 
yrs or longer, for a birth rate of 10$^{-2}$ yr$^{-1}$ or lower. The much 
smaller number of the AXPs may be due to the depletion of the supply of 
mass inflow in most sources by the end of the propeller phase. This
waiting time effect will be enhanced if the propeller phase is preceded
with an isolated pulsar (ejector) phase during which 
pulsar radiation prohibits gravitational capture and
mass inflow. In this phase the pulsar spins down rapidly through magnetic 
dipole radiation. The propeller phase starts when gravitational capture
of matter is permitted, that is, when the Shvartsman radius$^{50}$ at which 
the pulsar luminosity would stop mass inflow becomes smaller than the 
Bondi-Hoyle gravitational capture radius. This gives
\begin{equation}
\dot{M} = 2/3 \mu^2 \Omega^4 / C_s c^4
= 2 \times 10^{10} {\mu_{30}}^2 \Omega^4 / (kT (keV))^{1/2}
\end{equation}
where c is the speed of light, C$_s$ and T are the sound speed and 
temperature in the ambient medium. The initial phase of dipole
spindown would last until the critical period given by Eq.(14)
\begin{equation}
P = 0.47 {\mu_{30}}^{1/2} (kT (keV))^{-1/8} {{\dot{M}}_{15}}^{-1/4}
\end{equation}
is reached, at the time
\begin{equation}
t_1 = 3.5 \times 10^6 yrs I_{45} (kT (keV))^{-1/4}
{{\dot{M}}_{15}}^{-1/2} {\mu_{30}}^{-1}.
\end{equation}
Values of t$_1$ are given in Table 1. If the mass inflow is depleted
before a time of the order of t$_1$, if relevant, plus the duration of the 
propeller phase, which is of the
order of t$_{min}$, the subsequent accretion phases never occur. 
This may be the reason why the numbers of AXPs and SGRs are much less
than the number of DTNs. Further, one can understand, by the same 
token, why mass inflow rates $<$ 10$^{15}$ gm s$^{-1}$ are not inferred:  
at lower mass inflow rates the waiting times t$_1$ before the propeller
phase, and t$_{min}$ or longer during the propeller phase exceed the
lifetime of the mass inflow source. The rare AXPs
and SGRs are thus the young objects born with conventional magnetic
fields but in circumstances of large enough mass inflow towards the neutron
star, so that the timescales t$_1$ and t$_0$, which scale with
$\dot{M}^{-1/2}$ and $\dot{M}^{-1}$ respectively, are short compared to
the lifetime of the mass supply: 
These are the rare sources which have evaded the ejector phase and/or gone 
through the propeller phase in time to start accretion 
before the matter supply is
depleted. The more common source is the DTN, outliving an ejector phase and
now in a propeller phase of duration 10$^6$ yrs. If DTNs are born at
rates comparable to that of radio pulsars in supernova explosions, then a
significant fraction of  SNRs must leave the neutron star under mass
inflow conditions of $\dot{\em M}$ = 10$^{15}$ gm/s. For most DTNs the
propeller stage must be long enough that the source of mass inflow is
depleted before the start of accretion. 
(ii) Alternatively, the spindown could be taking place from a very low 
mass binary companion. At present 
the timing observations give a$_x$ sin i values which limit the companion 
mass to be less than a few tenths of a solar mass$^{16,29}$. 
There are serious constraints on this 
scenario: If all AXPs are to be descendants of DTNs, either the 
propeller phase lasts 1000 times longer than the AXP accretion phase, or only 
10$^{-3}$ of DTNs turn into AXPs, according to the comparison of total 
numbers of these two types of sources in the Galaxy. The DTN-propeller phase 
should start when the low mass companion evolves, to start mass inflow 
towards the neutron star. This will take longer than 10$^8$ yrs, followed by 
10$^6$ yrs for the DTN-propeller phase, and then a much shorter 
time, 10$^3$ yrs, for the AXP accretion phase. This is difficult to 
explain if the timescale of the mass inflow is the evolutionary timescale 
of the low mass binary. A comparison with the population of LMXBs with 
the assumption that a fraction f of LMXBs go through  the AXP phase gives 
f $t_{AXP} = (N_{AXP}/N_{LMXB}) t_{LMXB} \leq 
0.1 t_{LMXB} \sim $ 10$^7$ - 10$^8$ yrs, contradicting the estimate of 
$t_{AXP}$ = 10$^3$yrs from the comparison with DTNs. A way out of these 
difficulties within the LMXB scenario would be for the AXPs to represent a 
rare path in LMXB evolution and a low probability to evolve out of the 
propeller (DTN) stage. Taking into account also the evidence for possible 
association with supernova remnants for two AXPs, mass supply from a debris 
disk around a single neutron star seems to be the more likely scenario. 

The SGRs are like the AXPs in their X-ray properties. Both classes of
sources are detectable throughout the galaxy, and their comparable
numbers suggest the SGRs are also in the same rare or relatively short
beginning accretor phase as the AXPs. The associations of most
SGRs with SNRs and plerions, and their gamma-ray bursting activity
suggest that the SGRs are younger than the AXPs. While the present 
work does not address the mechanisms of the gamma ray
burst phenomenon, it is intriguing
that as Wang and Robertson$^{71}$ have noted propellers (and probably their
descendants the early accretors) can support relativistic particle 
luminosities and gamma ray production in the surrounding accumulated matter. 
The energy source could be sporadic accretion of accumulated circumstellar 
mass released through instabilities. 

{\bf Comptonized spectra: higher $\dot{M}$ sources may not allow pulsar beams}

 Why is it that the AXPs are observed as pulsars while the LMXBs are not 
(with the one important recently discovered exception XTE J1808-369 $^{55}$)? 
The proposal that millisecond radio pulsars have been spun-up 
through accretion in LMXBs$^{56,57}$ and the beat frequency model$^{58}$ 
which implied millisecond rotation periods for the neutron star 
in connection with the first discovered$^{59}$ quasi-periodic oscillations 
in LMXBs, were followed by extensive searches for millisecond pulsars in 
LMXBs all with null results$^{60-63}$.
The explanation$^{64}$ for the rarity of LMXB 
pulsars has been that comptonization by material around the neutron 
star will destroy pulses by washing out the beams emerging from 
the neutron star if the material is optically thick to electron 
scattering$^{65-68}$. The present picture is consistent with this: the AXPs 
are observed as pulsars because they are in the beginning stages of 
accretion, and the corona around them is yet to build up significant optical 
thickness to destroy the beaming. 

The power law photon energy spectra of AXPs are characterized by 
large photon indices (soft and steep spectra). The values of the photon 
index $\Gamma$ for AXPs range between 3  and 4, with only one source 
having an index of 2.5, while high mass x-ray binaries typically have 
power law indices of 1-2; among LMXBs with power law spectral fits, most 
have indices less than 2, with the highest value of 2.8 $^{69}$. 
According to the extensive simulations of
unsaturated comptonization by Pozdnyakov, Sobol and Sunyaev$^{70}$, 
the photon index is related to the 
compton y parameter or to a related parameter $\gamma$ 
through the expression: 
\begin{equation} 
\Gamma =  -1/2 + [ 9/4 + \gamma ]^{1/2}
\end{equation}
where fits to Monte Carlo simulations give: 
\begin{equation}
\gamma =  (\pi^2 / 4) (m_{el}c^2 /kT ) (\tau + 1/2)^{-2}
\end{equation}
for a comptonizing medium of optical thickness $\tau$ and spherical
symmetry. 
In the spectra of the AXPs there is no evidence 
for a high energy cut-off indicating the electron temperature. 
Taking the electron temperature to be at higher energies than the 
observational band, that is, taking kT $>$ 10 keV,   
limits on the optical thickness can be obtained. The models of Wang and 
Robertson$^{71}$ for the propeller stage give a temperature 35 keV 
in the plasma accumulated above the boundary of the magnetosphere, 
for P = 8 s, $\dot{M}$ = 10$^{15}$ gm/s, and allowed values of the 
dimensionless parameters $\eta $= 0.5, $\beta$ =0.01, and $\zeta$ = 1. 
With kT = 35 keV, the photon indices give optical thickness values of 
the order of one. These values are given in Table 2. 
One can use the estimates of Wang and Robertson for the 
density, of the order of 10$^{-9}$ gm/s accumulated at the boundary of 
the propeller 
magnetosphere, which gives an optical thickness to electron scattering 
$\tau \sim $1 for scattering medium dimensions of the order of 
r$_A \sim $r$_c \sim$ 10$^9$ cm in the systems of interest. 
The intrinsic luminosity can be estimated to be 0.2-0.3 of  
the luminosity emerging from the comptonized cloud using the feedback 
model of comptonized spectra$^{72}$. The 
intrinsic thermal Bremsstrahlung 
luminosity that Wang and Robertson have adopted for the plasma envelope 
can indeed amount to 0.2-0.3 of the observed luminosities. 
Whether there is an evolutionary connection or not, what distinguishes 
SGRs and AXPs from LMXBs is that as young accretors their 
circumstellar material is not optically thick to electron scattering yet. 
Thus the unsaturated comptonization spectra are soft and the beaming of 
the neutron star luminosity is not washed out by comptonization. 

{\bf On the unique millisecond x-ray pulsar SAX J1808.4-3658}

The recently discovered$^{55}$ source SAX J1808.4-3658 with a 2.5 ms period is 
likely to 
be an old LMXB if millisecond rotation powered pulsars evolve through  
spin up by accretion in old LMXBs that have weak magnetic fields$^{56,57}$. 
This is the only LMXB whose rotation period has been observed. 
SAX J1808.4-3658 can be explained qualitatively again by the 
properties of the comptonizing medium, this time in the 
opposite limit of very large relative contribution of intrinsic 
luminosity from the medium coupled with efficient feedback from the 
medium to the neutron star's surface luminosity. A first report of the 
spectrum noted a power law spectrum extending without break to energies
above 120 keV $^{73}$, suggesting an electron temperature $>$ 
120 keV for the comptonizing medium. Heindl and Smith $^{74}$ have examined 
the 2.5-250 keV spectrum with the PCA and HEXTE detectors on RXTE. 
They find a soft excess for which there is an ambiguity among fits with 
various spectral models, with a Comptonized model for the spectrum 
below 20 keV yielding kT = 22 keV and optical thickness $\tau$ = 4. 
When the broad band spectrum was fitted with an iron line + disk blackbody to 
cover the soft excess below 20 kev, plus a power law with exponential cutoff 
above 34 keV, with an e-folding energy (equivalent kT) of 127 keV, 
the power law index defining the spectrum at 20-150 keV was found to be 1.86. 
Gilfanov et al$^{75}$ have also analyzed the RXTE spectra in terms of 
Comptonized models, finding power law indices between 1.86 and 2.29 for 
different observations. They find that a cut-off energy must be at least 
100 keV, and the e-folding energy is at least 270 keV, 
and consider comptonization by the bulk motion of plasma 
surrounding this burster. 
Assuming the properties of the Comptonizing material determine the spectrum 
at the higher energies, above 20 keV, while the soft excess below 20 keV 
is dominated by the input source spectrum, we adopt a photon index 
$\Gamma$ =2, and an equivalent kT= 120 keV to describe the thermal and 
kinetic energy of bulk motion. This yields an optical 
thickness of 1.1 to electron scattering (Table 2), consistent with 
the explanation that this unique source 
of X-ray pulsations has an observable rotation period because unsaturated 
Comptonization ($\tau \sim$ 1) allows the beamed radiation to emerge without 
smearing. Gilfanov et al$^{75}$ have also noted that the luminosity 
decayed abruptly by a factor of about 30 within 5 days but the spectral 
characteristics, in particular the broad band power law photon index of 
$\Gamma \sim$ 2 did not change through this transition. They interpret the 
drop in luminosity as a transition from the accretion regime to the propeller 
regime. The implied constancy of comptonizing material around the star through 
transitions between accretion and propeller phases is in line with 
the assumption of the present model that the dynamical interaction 
with the circumstellar material (the torques) are continuous through the 
accretor-propeller transition. Transitions between propeller and accretion 
stages have been discussed recently also in connection with the 
X-ray transients Aquila X-1 $^{76,77}$. 
 
{\bf Predictions and summary of the model}

For both the DTNs and the AXPs there is another source of
energy dissipation in the disk or circumstellar material, due to
accretion down to $r_A$:
\begin{equation}
L(r_A) = 3/2 GM {\dot{M}} / r_A \sim 3/2 (GM)^{2/3}\Omega^{2/3}
{\dot{M}} \sim 1.3 \times 10^{33} erg s^{-1}
{\dot{M}}_{15} m^{2/3} P^{-2/3}
\end{equation}
with the corresponding effective temperature
\begin{equation}
T(r_A) = (L(r_A)/4 \pi {r_A}^2 \sigma)^{1/4}
\sim 9.5 \times 10^4 K  {{\dot{M}}_{15}}^{1/4} P^{-1/2}
\end{equation}
For the AXPs this luminosity is smaller than the accretion
luminosity by a factor
\begin{equation}
L(r_A)/L  \sim 3/2 R/r_A (\dot{\em M}/\dot{\em M}_{acc}) 
\sim 10^{-2} (\dot{\em M}/\dot{\em M}_{acc}) m^{-1/3} P^{-2/3}
\end{equation}
noting that the mass inflow may not be all accreted even in the AXPs. 
For the DTNs L(r$_A$) is actually larger than the observed luminosity
$\dot{E}_{diss}$ by a factor
\begin{equation}
L(r_A)/\dot{E}_{diss}  \sim (I \Omega |\dot\Omega|)/
(I_p \omega |\dot\Omega|)  \sim (10^{2}-10^{4}) \Omega.
\end{equation}
For both AXPs and DTNs the temperature in Eq.(20) applies, indicating 
the euv range for the radiation from the disk. This would be extremely 
difficult to detect, but being nearby sources, DTNs with low neutral 
hydrogen column density might provide a chance of looking for a disk 
luminosity as a test of the present model.

Thus a unified picture to include SGRs, DTNs and AXPs 
can be proposed. The salient features of this picture are: 
(i) The similarity in the rotation periods of these three 
classes of sources is not a coincidence but rather a consequence of 
the asymptotic approach to rotational equilibrium under similar circumstances. 
(ii) The observability of the rotation period in the AXPs but not in the 
LMXBs can be explained qualitatively in terms of comptonization as 
supported by an interpretation of their spectra.   
(iii) The circumstances are similar for the sources with observed rotation 
periods because of selection effects for $\dot{\em M}$. Larger 
$\dot{\em M}$ 
give optically thick comptonizing matter which suppresses the beaming of 
the radiation and renders the pulses at the rotation period of the star 
unobservable, while with low $\dot{\em M}$ the spindown is too slow 
for the source to go through the propeller stage and reach the accretion 
stage before the circumstellar mass is depleted. 
(iv) The DTNs are the first observed examples of neutron stars in the 
propeller phase. 
(v) A non-accreting neutron star under propeller spindown has a 
luminosity provided by energy dissipation inside the star. 
(vi) The SGRs are neutron stars accreting 
from the debris of their supernova remnants, from a leftover or 
protoplanetary disk.
(vi) The AXPs are also likely to be in an early stage of accretion, 
following the SGR stage. The source of the accreted mass is likely to 
be a finite store of debris around the neutron star, like a 
remnant disk. Alternatively there is a small chance that they 
might represent the beginning stages of a rare path in LMXB evolution. 
(vii) All these sources have conventional 
10$^{12}$ G magnetic fields, supporting the viewpoint that neutron stars 
are born with 10$^{12}$ G fields and the weak fields of the millisecond and 
binary pulsars and of the LMXBs that exhibit QPOs result from field decay on 
evolutionary timescales of the LMXBs (induced by spindown of the neutron star 
and/or by accretion). 

This picture obviates the need to postulate magnetars on the grounds 
that isolated neutron stars with ordinary 10$^{12}$ G fields 
cannot have spun down to $\sim$ 10 s periods within the estimated ages. 
Whether the 
large spindown rates observed are due to spindown by interaction 
with ambient matter, as proposed here, or to spindown by a magnetar 
can be decided by detailed analysis of the fluctuations (noise) in the 
spindown process, as the timing noise characteristic of accretion 
powered neutron stars is quite distinguishable from the timing noise in 
the typically much quieter isolated rotation powered pulsars. This 
analysis will require frequently sampled timing observations of the 
AXPs and SGRs. Observation of a spindown rate in the expected range from 
RXJ0720.4-3125 would constitute strong evidence for the propeller hypothesis.  
The expected euv radiation from the disk is another prediction that could 
be checked in the nearby DTNs, though absorption in the euv band makes 
this difficult.

I thank H. \"Ogelman, \c{S}.Balman, A.Baykal, A.Esendemir, O. Guseinov, 
\"U.K{\i}z{\i}lo\u{g}lu and members of the High Energy Astrophysics Research 
Unit at METU for useful comments, the Scientific and Technical Research 
Council of Turkey (T\"UB\.{I}TAK) and the Turkish Academy of Sciences for 
partial support.

\newpage
\noindent
{\bf References}

\noindent
1. Mereghetti,S.\& Stella,L.The very low mass x-ray binary pulsars.{\em
Astrophys.J.}{\bf 442},L17-L20(1995).\\
2. van Paradijs,J.,Taam, R.E.\& van den Heuvel,E.P.J.On the nature of
the 'anomalous'6-s x-ray pulsars.{\em Astron.Astrophys.}
{\bf 299},L41-45(1995).\\
3. Ghosh,P., Angelini,L.\& White,N.E.The nature of the "6 second" and
related x-ray pulsars: evolutionary and dynamical considerations.
{\em Astrophys.J.}{\bf 478},713-722(1997).\\
4. Corbet,R.H.D., Smale,A.P.,Ozaki, M., Koyama,K. \& Iwasawa,K.The
spectrum and pulses of 1E2259+586 from ASCA and BBXRT observations.
{\em Astrophys.J.}{\bf 443},786-794(1995).\\
5. Iwasawa,K., Koyama,K. \& Halpern,J.P.Pulse period history and
cyclotron resonance feature of the X-ray pulsar 1E2259+586. {\em Publ.
Astron. Soc. Japan}{\bf 44},9-14(1992).\\
6. Baykal,A. \& Swank,J. Pulse frequency changes of 1E2259+586 and the
binary interpretation. {\em Astrophys.J.}{\bf 460},470-477(1996).\\
7. Parmar,A.N. {\em et al.} A BeppoSAX observation of the
X-ray pulsar 1E2259+586 and the supernova remnant G109.1-1.0 (CTB109).
{\em Astron.Astrophys.}{\bf 330},175-180(1998).\\
8. Mereghetti,S. A spin-down variation in the 6s X-ray pulsar
1E1048.1-5937. {\em Astrophys.J.}{\bf 455},598-(1995).\\
9. Corbet,R.H.D. \& Mihara,T. The spin-down rate and X-ray flux of
1E1048.1-5937. {\em Astrophys.J.}{\bf 475},L127-L130(1997).\\
10. Oosterbroek,T., Parmar,A.N., Mereghetti,S. \& Israel,G.L. The two
component X-ray spectrum of the 6.4s pulsar 1E1048.1-5937.
{\em Astron.Astrophys.}{\bf 334},925-930(1998).\\
11. White,N.E., Angelini,L., Ebisawa,K., Tanaka,Y. \& Ghosh,P.
The spectrum of the 8.7s X-ray pulsar 4U0142+161.
{\em Astrophys.J.}{\bf 463}, L83-L86(1996).\\
12. Hellier,C. A ROSAT observation of the X-ray pulsars X0142+614 and
X0146+612. {\em Mon. Not. Royal Astron. Soc.}{\bf 271},L21-L24(1994).\\
13. Sugizaki,M. {\em et al.} Discovery of an 11-s X-ray pulsar
in the galactic plane section of the Scorpius constellation.
{\em Publ. Astron. Soc. Japan}{\bf 49}, L25-L30(1997).\\
14. Vasisht,G. \& Gotthelf,E.V. The discovery of an anomalous X-ray
pulsar in the supernova remnant Kes 73. 
{\em Astrophys.J.}{\bf 486},L129-L132(1997).\\
15. Gotthelf,E.V. \& Vasisht,G. Discovery of a 7 second anomalous X-ray
pulsar in the distant Milky Way.{\em New Astronomy}{\bf 3},293-300(1998).\\
16. Mereghetti,S., Stella,L. \& Israel,G.L. Recent results on the
anomalous X-ray pulsars. Preprint, astro-ph 9712254 (1997).\\
17. Walter,F.M.,Wolk,S.J.\& Neuhauser,R.Discovery of a nearby isolated
neutron star.{\em Nature}{\bf 379},233-235(1996).\\
18. Walter,F.M.\&Matthews,L.D.The optical counterpart of the isolated
neutron star RXJ185635-3754.{\em Nature}{\bf 389},358-360(1997).\\
19. Haberl,F.,Motch,C.,Buckley,D.A.H.,Zickgraf,F.-J.\& Pietsch,W.
RXJ0720.4-3125: strong evidence of an isolated pulsating neutron star.
{\em Astron.Astrophys.}{\bf 326},662-668(1997).\\
20. Motch,C.\& Haberl,F. Constraints on optical emission from the
isolated neutron star candidate RXJ0720.4-3125. astro-ph-9809140,
submitted to {\em Astron.Astrophys.}(1998).\\
21. Kulkarni,S. \& van Keerkwijk,M.H. Optical observations of the
isolated neutron star RXJ0720.4-3125. preprint (1998).\\
22. Schwope, A.D., Hasinger,G., Schwarz,R., Haberl,F. \& Schmidt,M. The
isolated neutron star candidate RBS1223 (1RXSJ130848.6+212708). 
astro-ph-9811326, submitted to {\em Astron.Astrophys.} (1998).\\
23. Haberl,F., Motch,C. \& Pietsch,W. Isolated neutron stars in the 
ROSAT survey. {\em AN}{\bf 319},97(1998).\\
24. Kouveliotou,C. {\em et al.} An x-ray pulsar with a superstrong magnetic 
field in the soft $\gamma$-ray repeater SGR1806-20.{\em Nature}{\bf 
393},235-237(1998).\\
25. Hurley,K. {\em et al.}; Kouveliotou,C. {\em et al.}{\em IAU Circ.}
No.7001(1998).\\
26. Cline,T.L., Mazets,E.P. \& Golenetskii,S.V.{\em IAU Circ.}No.7002(1998).\\
27. Hurley,K. {\em et al.}ASCA discovery of an X-ray pulsar in the error box 
of SGR1900+14.astro-ph-9811388(1998).\\
28. Hurley,K. {\em et al.} A giant, periodic flare from the soft gamma 
repeater SGR1900+14.astro-ph-9811443(1998).\\
29. Kouveliotou,C. {\em et al.} Discovery of a magnetar associated with the 
soft gamma repeater SGR 1900+14. astro-ph-9809140(1998).\\
30. Thompson,C. \& Duncan,R.C. Neutron star dynamos and the origin of
pulsar magnetism.{\em Astrophys.J.}{\bf 408},194-217(1993).\\
31. Duncan,R.C. \& Thompson,C. Formation of very strongly magnetized
neutron stars: implications for gamma-ray bursts.{\em Astrophys.J.}{\bf
392},L9-L13 (1992).\\
32. Thompson,C. \& Duncan,R.C. The soft gamma repeaters as very strongly
magnetized neutron stars - I.Radiative mechanism for outbursts. {\em
Mon. Not. Royal Astron. Soc.}{\bf 275},255-300(1995).\\
33. Thompson,C. \& Duncan,R.C. The soft gamma repeaters as very strongly
magnetized neutron stars - II.Quiescent neutrino, x-ray and Alfven wave
emission. {\em Astrophys.J.}{\bf 473},322-342(1996).\\
34. Heyl,J.S. \& Hernquist,L. What is the nature of RXJ0720.4-3125?
preprint, astro-ph 9801038 (1998).\\
35. \"Ogelman,H. X-ray observations of cooling neutron stars, in
{\em The Lives of the Neutron Stars}(eds Alpar,M.A.,
K{\i}z{\i}lo\u{g}lu,\"U. \& van Paradijs,J.)101-120, 
(NATO ASI Series C{\bf 450}, Kluwer Academic Publishers, 1995).\\
36. Tsuruta,S. Thermal evolution of neutron stars: current status. in
{\em The Lives of the Neutron Stars}(eds Alpar,M.A.,
K{\i}z{\i}lo\u{g}lu,\"U. \& van Paradijs,J.)133-146, 
(NATO ASI Series C{\bf 450}, Kluwer Academic Publishers, 1995).\\
37. Page,D. Thermal evolution of isolated neutron stars, in
{\em The Many Faces of Neutron Stars}(eds Buccheri,R., van Paradijs,J.
\& Alpar,M.A.)539-551, (NATO ASI Series C{\bf 515}, Kluwer Academic Publishers, 1995).\\
38. Alpar,M.A., Anderson,P.W., Pines,D. \& Shaham,J. Vortex creep and
the internal temperature of neutron stars. I.General theory.
{\em Astrophys. J.}{\bf 276},325-334(1984).
39. Alpar,M.A., Nandkumar,R. \& Pines,D. Vortex creep and the internal
temperature of neutron stars: The Crab pulsar and PSR 0525+21.
{\em Astrophys. J.}{\bf 288},191-195(1985).\\
40. Shibazaki,N. \& Lamb,F.K. Neutron star evolution with internal
heating. {\em Astrophys. J.}{\bf 346},808-822(1989).\\
41. Umeda,H., Shibazaki,N., Nomoto,K. \& Tsuruta,S. Thermal evolution of
neutron stars with internal frictional heating.
{\em Astrophys. J.}{\bf 408},186-193(1993).\\
42. Becker,W. \& Tr\"umper,J. The x-ray luminosity of rotation powered
neutron stars.{\em Astron. Astrophys.}{\bf 326}, 682-691(1997).\\
43. Alpar,M.A., Brinkmann,W., K{\i}z{\i}lo\u{g}lu, \"U., \"Ogelman, H. \&
Pines, D., A search for X-ray emission from a nearby pulsar: PSR
1929+10. {\em Astron. Astrophys.}{\bf 276}, 101-104(1987).\\
44. Yancopoulos,S., Hamilton,T.T. \& Helfand,D.J. The detection of
pulsed X-ray emission from a nearby radio pulsar. {\em Astrophys. J.}{\bf 429},832-843(1994).\\
45. Alpar,M.A. Glitch dynamics and energy dissipation in neutron stars,
in {\em Neutron Stars and Pulsars} (eds Shibazaki,N., Kawai,N.,
Shibata,S. \& Kifune,T.)129-138. (Universal Academic Publishers, 
Tokyo, 1998).\\ 
46. Shemar,S.L. \& Lyne,A.G. Observations of pulsar glitches. 
{\em Mon. Not. Royal Astron. Soc.}{\bf 282},677-690(1996).\\
47. Alpar,M.A. \& Baykal,A. Expectancy of large pulsar glitches: a
comparison of models with the observed glitch sample. {\em Mon. Not.
Royal Astron. Soc.}{\bf 269},849-856(1994).\\
48. Alpar,M.A., Chau,H.F., Cheng,K.S. \& Pines,D. Postglitch Relaxation
of the Vela pulsar after its first eight large glitches. {\em Astrophys. 
J.}{\bf 409},345-359(1993).\\
49. Illarionov,A.F. \& Sunyaev,R.A. Why the number of galactic X-ray
stars is so small. {\em Astron. Astrophys.}{\bf 39},185-195(1975).\\
50. Lipunov,V.M. {\em Astrophysics of Neutron Stars}, Springer(1992).\\
51. Frank,J., King,A. \& Raine,D. {\em Accretion Power in Astrophysics}, 
Cambridge University Press(1992).\\
52. Ghosh,P. \& Lamb,F.K. Accretion by rotating magnetic neutron stars.
III. Accretion torques and period changes in pulsating X-ray sources. 
{\em Astrophys. J.}{\bf 234},296-316(1979).\\
53. Ghosh,P. \& Lamb,F.K. Plasma physics of accreting neutron stars, in
{\em Neutron Stars: Theory and Observation}(eds Ventura,J. \& Pines,D.)363-444,
NATO ASI Series C{\bf 344}, Kluwer Academic Publishers, 1991).\\
54. Srinivasan,G., Bhattacharya,D., Muslimov,A.G. \& Tsygan,A.I. 
A novel mechanism for the decay of neutron star magnetic fields.	 
{\em Curr. Sci.}{\bf 59},31-38(1990).\\
55. Wijnands,R. \& van der Klis,M. A millisecond pulsar in an X-ray 
binary system. {\em Nature}{\bf 394},344-346(1998).\\ 
56. Alpar,M.A., Cheng,A.F., Ruderman,M.A. \& Shaham,J. A new class of
radio pulsars. {\em Nature}{\bf 300},728-730(1982).\\
57. Radhakrishnan,V. \& Srinivasan,G. on the origin of the recently
discovered ultra-rapid pulsar. {\em Curr. Sci.}{\bf 51},1096-1099(1982).\\
58. Alpar,M.A. \& Shaham,J. Is GX 5-1 a millisecond pulsar? 
{\em Nature}{\bf 316},239-241(1985).\\
59. van der Klis,M. {\em et al.} Intensity-dependent quasi-periodic 
oscillations in the X-ray flux of GX 5-1.{\em Nature}{\bf 316},225-230(1985).\\
60. Leahy,D.A. {\em et al.} On searches for pulsed emission with 
application to four globular cluster x-ray sources. 
{\em Astrophys. J.}{\bf 266},160-170(1983).\\ 
61. Mereghetti,S. \& Grindlay,J. A search for millisecond periodic and 
quasi-periodic pulsations in low mass x-ray binaries. {\em Astrophys. J.}
{\bf 312},727-731(1987).\\
62. Wood,K.S. {\em et al.} Searches for millisecond pulsations in low 
mass x-ray binaries. I. {\em Astrophys. J.}{\bf 379},295-309(1991).\\
63. Vaughan,B. {\em et al.} Searches for millisecond pulsations in low 
mass x-ray binaries. II. {\em Astrophys. J.}{\bf 435},362-371(1994).\\
64. Lamb,F.K., Shibazaki,N., Alpar,M.A. \& Shaham,J. Quasi-periodic 
oscillations in bright galactic-bulge X-ray sources. {\em Nature}{\bf 
317},681-687(1985).\\
65. Brainerd,J. \& Lamb,F.K. Effect of an electron scattering cloud on 
x-ray oscillations by beaming. {\em Astrophys. J.}{\bf 
317},L33-L38(1987).\\ 
66. Kylafis,N.D. \& Klimis,G.S. Effects of electron scattering on the 
oscillations of an x-ray source. {\em Astrophys. J.}{\bf 323},678-684(1987).\\ 
67. Wang,Y.-M. \& Schlickeiser,R. Smearing of a beaming pattern by an 
isotropic cloud: an analysis with applications to non-pulsing x-ray 
sources and cosmic rays. {\em Astrophys. J.}{\bf 313},200-217(1987).\\
68. Bussard,R.W., Weisskopf,M.C., Elsner,R.F. \& Shibazaki,N. The effect 
of a hot, spherical scattering cloud on quasi-periodic oscillation 
behavior. {\em Astrophys. J.}{\bf 327},284-293(1988).\\
69. Guseinov,O.H. et al. A catalogue of low and high mass X-ray binaries. 
http://astroa.physics.metu.edu.tr (1998).\\
70. Pozdnyakov,L.A., Sobol,I.M. \& Sunyaev,R.A. Comptonization and the 
shaping of X-ray source spectra: Monte Carlo calculations. {\em 
Astrophys. Space Phys. Rev.}{\bf 2},189-331(1983).\\
71. Wang,Y.-M. \& Robertson,J.A. "propeller" action by rotating neutron 
stars. {\em Astron. Astrophys.}{\bf 151},361-371(1985).\\ 
72. Gilfanov,M. et al. Hard X-ray observations of black hole candidates.
{\em The Lives of the Neutron Stars}(eds Alpar,M.A.,
K{\i}z{\i}lo\u{g}lu,\"U. \& van Paradijs,J.)331-354, 
(NATO ASI Series C{\bf 450}, Kluwer Academic Publishers, 1995).\\
73. Marshall,F.E. SAX J1808.4-3658 = XTE J1808-369. 
{\em IAU Circ.} No.6876(1998).\\   
74. Heindl,W.A. \& Smith,D.M. The x-ray spectrum of SAX J1808.4-3658. 
{\em Astrophys. J.}{\bf 506},L33-L38(1998).\\
75. Gilfanov,M., Revnintsev,M., Sunyaev,R. \& Churazov,E. The millisecond 
x-ray pulsar/burster SAX J1808.4-3658: the outburst light cuurve and the 
power spectrum. {\em Astron. Astrophys.}{\bf 338},L83-L86(1998).\\
76. Zhang,S.N., Yu,W. \& Zhang,W. Spectral state transitions in 
Aquila X-1: evidence for "propeller" effects. {\em Astrophys. J.}{\bf 494},
L71-L74(1998).\\
77. Campana,S. {\em et al.} Aquila X-1 from outburst to quiescence: the 
onset of the propeller effect and signs of a turned-on rotation powered 
pulsar. {\em Astrophys. J.}{\bf 499},L65-L68(1998).\\

\newpage
\vspace{0.5cm}
\begin{table}
\begin{center}
{\bf Table 1 Magnetic moments and timescales }
\begin{tabular}{cllllllll}
\hline
Source & P (s) & $\dot{P}$ (s s$^{-1}$) & $\dot{M}_{15}$ & P$_{eq}$ (s) &
t$_o$ (yrs)
 & $\mu_{30}$  & t$_{min}$ (yrs) & t$_1$ (yrs) \\
\hline
1E1048.1-5937 & 6.44 & 2 $\times$ 10$^{-11}$ & 1 & 23 & 7.4 $\times$ 10$^3$
 & 1.45  & 5 $\times$ 10$^4$ & 2.4 $\times$ 10$^6$ \\
              &      &                   & 5 & 10.8 & 4.1 $\times$ 10$^3$
 & 1.33 & 3 $\times$ 10$^4$ & 1.2 $\times$ 10$^6$ \\
1E2259+586 & 6.98 & 6 $\times$ 10$^{-13}$ & 2 & 7.4 & 1.7 $\times$ 10$^4$
 & 0.54 & 1.6 $\times$ 10$^5$ & 4.6 $\times$ 10$^6$ \\
           &      &                   & 100 & 6.99 & 367
 & 3.57 & 4.9 $\times$ 10$^3$ & 9.8 $\times$ 10$^4$ \\
4U0142+61 & 8.69 & 2.1 $\times$ 10$^{-12}$ & 1 & 10.5 & 2.1 $\times$ 10$^4$
 & 0.57 & 1.8 $\times$ 10$^5$ & 6.1 $\times$ 10$^6$ \\
\hline
SGR1806-20 & 7.47 & 8.3 $\times$ 10$^{-11}$ & 100 & 13.8 & 148
 & 7.9 & 1.1 $\times$ 10$^3$ & 4.4 $\times$ 10$^4$ \\
SGR1900+14 & 5.16  & 1.1 $\times$ 10$^{-10}$ & 20 & 9.1 & 1.3 $\times$ 10$^3$
 & 2.2 & 9.1 $\times$ 10$^3$ & 3.6 $\times$ 10$^5$ \\
\hline
RXJ0720.4-3125 & 8.39 & (2.9 $\times$ 10$^{-11}$) & 1 & 10.8 & 2 $\times$
10$^4$
 & 0.59 & 1.7 $\times$ 10$^5$ & 5.9 $\times$ 10$^6$ \\
\hline
\end{tabular}

\vspace{0.5cm}

For RXJ0720.4-3125 the $\dot{P}$ value is inferred from the luminosity 
according to Eq.(1).
\end{center}
\end{table}

\vspace{25cm}

\begin{table}
\begin{center}
{\bf Table 2 Photon indices and optical thickness to electron scattering}
\begin{tabular}{clll}
\hline
Source & $\Gamma$ & $\tau$(kT=10keV) & $\tau$(kT=35keV) \\
\hline
4U0142+61 & 3.67 & 2.4 & 1.04 \\
1E1048.1-5937 & 2.5 & 3.8 & 1.8 \\
1E1841-045 & 3 & 3.4 & 1.4 \\
RXJ170849.0-400910 & 2.9 & 3.6 & 1.47 \\
\hline
eg & 2 & 5.1 & 2.5 \\
eg & 1.5 & 8 & 4 \\
\hline
SAX J1808.4-3658 & 2.02 & $\tau$(kT=120 keV)&=1.1 \\
\hline
\end{tabular}

\vspace{0.5cm}

\end{center}
The rows labeled "eg" give examples for typical power law spectra 
for x-ray binaries.
For SAX J1808.4-3658 $\tau$ is quoted for an equivalent temperature of 120 keV.
\end{table}
\end{document}